\pdfoutput=1
\documentclass[12pt,a4paper]{article}
\usepackage{ifthen} 
\newboolean{pdflatex}
\setboolean{pdflatex}{true} 

\newboolean{articletitles}
\setboolean{articletitles}{true} 

\newboolean{uprightparticles}
\setboolean{uprightparticles}{false} 

\newboolean{inbibliography}
\setboolean{inbibliography}{false} 

\usepackage{microtype}
\usepackage{lineno}  
\usepackage{xspace} 
\usepackage{caption} 

\usepackage{graphicx}  
\usepackage{color}
\usepackage{colortbl}
\graphicspath{{./figs/}} 

\usepackage{amsmath} 
\usepackage{amssymb}
\usepackage{amsfonts}
\usepackage{upgreek} 

\newcommand*\patchAmsMathEnvironmentForLineno[1]{%
\expandafter\let\csname old#1\expandafter\endcsname\csname #1\endcsname
\expandafter\let\csname oldend#1\expandafter\endcsname\csname
end#1\endcsname
 \renewenvironment{#1}%
   {\linenomath\csname old#1\endcsname}%
   {\csname oldend#1\endcsname\endlinenomath}%
}
\newcommand*\patchBothAmsMathEnvironmentsForLineno[1]{%
  \patchAmsMathEnvironmentForLineno{#1}%
  \patchAmsMathEnvironmentForLineno{#1*}%
}
\AtBeginDocument{%
\patchBothAmsMathEnvironmentsForLineno{equation}%
\patchBothAmsMathEnvironmentsForLineno{align}%
\patchBothAmsMathEnvironmentsForLineno{flalign}%
\patchBothAmsMathEnvironmentsForLineno{alignat}%
\patchBothAmsMathEnvironmentsForLineno{gather}%
\patchBothAmsMathEnvironmentsForLineno{multline}%
}

\usepackage{hyperref}    
\usepackage[all]{hypcap} 

\def\lhcb {\mbox{LHCb}\xspace}

\def\cms    {\mbox{CMS}\xspace}

\def\cdf    {\mbox{CDF}\xspace}

\def\lhc    {\mbox{LHC}\xspace}

\ifthenelse{\boolean{uprightparticles}}%
{

 \def\Pmu         {\ensuremath{\upmu}\xspace}

 \def\Ppsi        {\ensuremath{\uppsi}\xspace}

 \def\PDelta      {\ensuremath{\Delta}\xspace}                 
 \def\PXi      {\ensuremath{\Xi}\xspace}                 
 \def\PLambda      {\ensuremath{\Lambda}\xspace}                 
 \def\PSigma      {\ensuremath{\Sigma}\xspace}                 
 \def\POmega      {\ensuremath{\Omega}\xspace}                 
 \def\PUpsilon      {\ensuremath{\Upsilon}\xspace}                 
 
 \def\PB      {\ensuremath{\mathrm{B}}\xspace}                 
                  
 \def\PD      {\ensuremath{\mathrm{D}}\xspace}

 \def\PJ      {\ensuremath{\mathrm{J}}\xspace}                 
 \def\PK      {\ensuremath{\mathrm{K}}\xspace}

 \def\Pb      {\ensuremath{\mathrm{b}}\xspace}                 
 \def\Pc      {\ensuremath{\mathrm{c}}\xspace}

 \def\Pi      {\ensuremath{\mathrm{i}}\xspace}

}
{

 \def\Pmu         {\ensuremath{\mu}\xspace}

 \def\Ppsi        {\ensuremath{\psi}\xspace}                 
                  
 \mathchardef\PDelta="7101
 \mathchardef\PXi="7104
 \mathchardef\PLambda="7103
 \mathchardef\PSigma="7106
 \mathchardef\POmega="710A
 \mathchardef\PUpsilon="7107
                  
 \def\PB      {\ensuremath{B}\xspace}                 
                  
 \def\PD      {\ensuremath{D}\xspace}

 \def\PJ      {\ensuremath{J}\xspace}                 
 \def\PK      {\ensuremath{K}\xspace}

 \def\Pb      {\ensuremath{b}\xspace}                 
 \def\Pc      {\ensuremath{c}\xspace}

 \def\Pi      {\ensuremath{i}\xspace}

}

\def\mup        {\ensuremath{\Pmu^+}\xspace}
\def\mun        {\ensuremath{\Pmu^-}\xspace} 
\def\mumu       {\ensuremath{\Pmu^+\Pmu^-}\xspace}

\def\cquark    {\ensuremath{\Pc}\xspace}

\def\bquark    {\ensuremath{\Pb}\xspace}

\def\kaon  {\ensuremath{\PK}\xspace}
  \def\Kbar  {\kern 0.2em\overline{\kern -0.2em \PK}{}\xspace}

\def\Kp    {\ensuremath{\kaon^+}\xspace}

  \def\Dbar    {\kern 0.2em\overline{\kern -0.2em \PD}{}\xspace}

\def\B       {\ensuremath{\PB}\xspace}
\def\Bbar    {\ensuremath{\kern 0.18em\overline{\kern -0.18em \PB}{}}\xspace}

\def\Bu      {\ensuremath{\B^+}\xspace}

\def\Bp      {\ensuremath{\Bu}\xspace}

\def\jpsi     {\ensuremath{{\PJ\mskip -3mu/\mskip -2mu\Ppsi\mskip 2mu}}\xspace}
\def\psitwos  {\ensuremath{\Ppsi{(2S)}}\xspace}

  \def\Y#1S{\ensuremath{\PUpsilon{(#1S)}}\xspace}

\def\Lbar {\ensuremath{\kern 0.1em\overline{\kern -0.1em\PLambda}}\xspace}

\newcommand{\decay}[2]{\ensuremath{#1\!\to #2}\xspace}         

\def\to                 {\ensuremath{\rightarrow}\xspace}

\def\AT#1     {\ensuremath{A_{\mathrm{T}}^{#1}}\xspace}           

\def\C#1      {\ensuremath{\mathcal{C}_{#1}}\xspace}                       
\def\Cp#1     {\ensuremath{\mathcal{C}_{#1}^{'}}\xspace}                    
\def\Ceff#1   {\ensuremath{\mathcal{C}_{#1}^{\mathrm{(eff)}}}\xspace}        
\def\Cpeff#1  {\ensuremath{\mathcal{C}_{#1}^{'\mathrm{(eff)}}}\xspace}       
\def\Ope#1    {\ensuremath{\mathcal{O}_{#1}}\xspace}                       
\def\Opep#1   {\ensuremath{\mathcal{O}_{#1}^{'}}\xspace}                    

\newcommand{\tev}{\ifthenelse{\boolean{inbibliography}}{\ensuremath{~T\kern -0.05em eV}\xspace}{\ensuremath{\mathrm{\,Te\kern -0.1em V}}\xspace}}
\newcommand{\gev}{\ensuremath{\mathrm{\,Ge\kern -0.1em V}}\xspace}
\newcommand{\mev}{\ensuremath{\mathrm{\,Me\kern -0.1em V}}\xspace}
\newcommand{\kev}{\ensuremath{\mathrm{\,ke\kern -0.1em V}}\xspace}
\newcommand{\ev}{\ensuremath{\mathrm{\,e\kern -0.1em V}}\xspace}
\newcommand{\gevc}{\ensuremath{{\mathrm{\,Ge\kern -0.1em V\!/}c}}\xspace}
\newcommand{\mevc}{\ensuremath{{\mathrm{\,Me\kern -0.1em V\!/}c}}\xspace}
\newcommand{\gevcc}{\ensuremath{{\mathrm{\,Ge\kern -0.1em V\!/}c^2}}\xspace}
\newcommand{\gevgevcccc}{\ensuremath{{\mathrm{\,Ge\kern -0.1em V^2\!/}c^4}}\xspace}
\newcommand{\mevcc}{\ensuremath{{\mathrm{\,Me\kern -0.1em V\!/}c^2}}\xspace}

\def\mum  {\ensuremath{{\,\upmu\rm m}}\xspace}

\def\invfb   {\ensuremath{\mbox{\,fb}^{-1}}\xspace}

\newcommand{\chisq}{\ensuremath{\chi^2}\xspace}

\def\gsim{{~\raise.15em\hbox{$>$}\kern-.85em
          \lower.35em\hbox{$\sim$}~}\xspace}
\def\lsim{{~\raise.15em\hbox{$<$}\kern-.85em
          \lower.35em\hbox{$\sim$}~}\xspace}

\def\sPlot{\mbox{\em sPlot}\xspace}
\def\sWeight{\mbox{\em sWeight}\xspace}

\def\sqs   {\ensuremath{\protect\sqrt{s}}\xspace}

\def\pt         {\mbox{$p_{\rm T}$}\xspace}

\def\evtgen     {\mbox{\textsc{EvtGen}}\xspace}

\def\geant      {\mbox{\textsc{Geant4}}\xspace}

\def\photos     {\mbox{\textsc{Photos}}\xspace}

\def\pythia     {\mbox{\textsc{Pythia}}\xspace}

\def\tell1  {TELL1\xspace}
\def\ukl1   {UKL1\xspace}

\usepackage{cite} 
\usepackage{mciteplus}

\newcommand{\gevct}{\ensuremath{{\,(\mathrm{Ge\kern -0.1em V\!/}c)^2}}\xspace}
\newcommand{\lla}{\ensuremath{\lambda_\theta}\xspace}
\newcommand{\llb}{\ensuremath{\lambda_{\theta\phi}}\xspace}
\newcommand{\llc}{\ensuremath{\lambda_{\phi}}\xspace}
\newcommand{\lli}{\ensuremath{\lambda_{\mathrm{inv}}}\xspace}

\usepackage{longtable} 
\usepackage{multirow} 
\usepackage{booktabs} 
\usepackage{lscape}

\begin{document}

\renewcommand{\thefootnote}{\fnsymbol{footnote}}
\setcounter{footnote}{1}


\begin{titlepage}
\pagenumbering{roman}

\vspace*{-1.5cm}
\centerline{\large EUROPEAN ORGANIZATION FOR NUCLEAR RESEARCH (CERN)}
\vspace*{1.5cm}
\hspace*{-0.5cm}
\begin{tabular*}{\linewidth}{lc@{\extracolsep{\fill}}r}
\ifthenelse{\boolean{pdflatex}}
{\vspace*{-2.7cm}\mbox{\!\!\!\includegraphics[width=.14\textwidth]{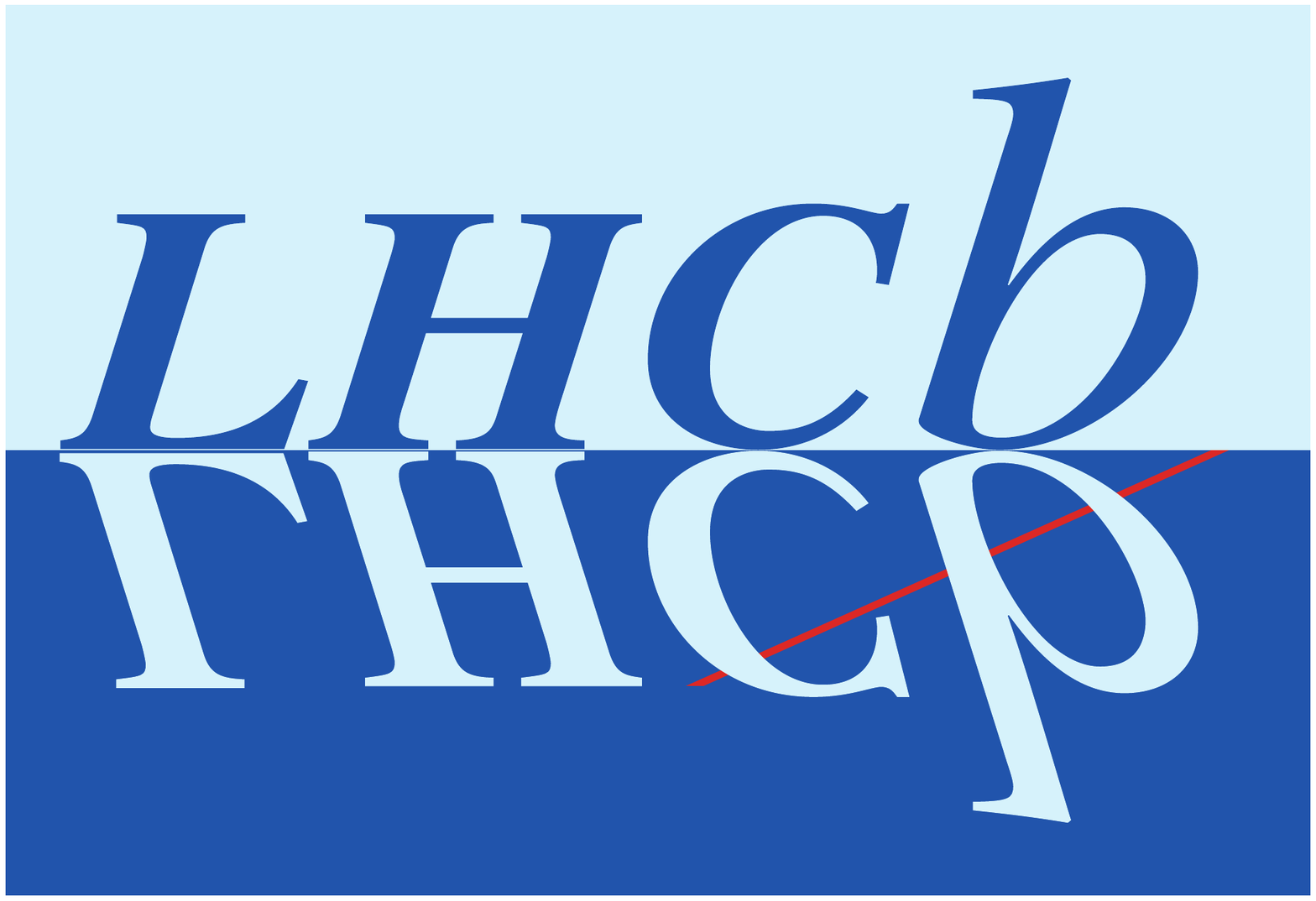}} & &}%
{\vspace*{-1.2cm}\mbox{\!\!\!\includegraphics[width=.12\textwidth]{lhcb-logo.eps}} & &}%
\\
 & & CERN-PH-EP-2014-029\\  
 & & LHCb-PAPER-2013-067\\  
 & & \today \\ 
\end{tabular*}

\vspace*{4.0cm}

{\bf\boldmath\huge
\begin{center}
   Measurement of $\psitwos$ polarisation 
   in $pp$ collisions at $\sqrt{s}=7\tev$
\end{center}
}

\vspace*{2.0cm}

\begin{center}
The LHCb collaboration\footnote{Authors are listed on the following pages.}
\end{center}

\vspace{\fill}

\begin{abstract}
  \noindent
The polarisation of prompt \psitwos mesons is measured by performing
an angular analysis of \decay{\psitwos}{\mumu} decays using proton-proton collision data,
corresponding to an integrated luminosity of 1.0\invfb, collected by the
LHCb detector at a centre-of-mass energy of $7 \tev$.  
The polarisation is measured in bins of
transverse momentum $\pt$ and rapidity $y$ in the kinematic region $3.5<\pt<15\gevc$ and 
$2.0<y<4.5$, and is compared to theoretical models.
No significant polarisation is observed. 

\end{abstract}

\vspace*{2.0cm}

\begin{center}
  Submitted to Eur.~Phys.~J.~C
\end{center}

\vspace{\fill}

{\footnotesize 
\centerline{\copyright~CERN on behalf of the \lhcb collaboration, license \href{http://creativecommons.org/licenses/by/3.0/}{CC-BY-3.0}.}}
\vspace*{2mm}

\end{titlepage}


\newpage
\setcounter{page}{2}
\mbox{~}
\newpage

\centerline{\large\bf LHCb collaboration}
\begin{flushleft}
\small
R.~Aaij$^{41}$, 
B.~Adeva$^{37}$, 
M.~Adinolfi$^{46}$, 
A.~Affolder$^{52}$, 
Z.~Ajaltouni$^{5}$, 
J.~Albrecht$^{9}$, 
F.~Alessio$^{38}$, 
M.~Alexander$^{51}$, 
S.~Ali$^{41}$, 
G.~Alkhazov$^{30}$
% ...

{\footnotesize \it
% ...
$ ^{5}$Clermont Universit\'{e}, Universit\'{e} Blaise Pascal, CNRS/IN2P3, LPC, Clermont-Ferrand, France\\
% ...
$ ^{9}$Fakult\"{a}t Physik, Technische Universit\"{a}t Dortmund, Dortmund, Germany\\
% ...
$ ^{30}$Petersburg Nuclear Physics Institute (PNPI), Gatchina, Russia\\
% ...
$ ^{37}$Universidad de Santiago de Compostela, Santiago de Compostela, Spain\\
$ ^{38}$European Organization for Nuclear Research (CERN), Geneva, Switzerland\\
% ...
$ ^{41}$Nikhef National Institute for Subatomic Physics, Amsterdam, The Netherlands\\
% ...
$ ^{46}$H.H. Wills Physics Laboratory, University of Bristol, Bristol, United Kingdom\\
% ...
$ ^{51}$School of Physics and Astronomy, University of Glasgow, Glasgow, United Kingdom\\
$ ^{52}$Oliver Lodge Laboratory, University of Liverpool, Liverpool, United Kingdom\\
% ...
}
\end{flushleft}

%
%

\cleardoublepage

\renewcommand{\thefootnote}{\arabic{footnote}}
\setcounter{footnote}{0}

\pagestyle{plain} 
\setcounter{page}{1}
\pagenumbering{arabic}

\section{Introduction}
\label{intro}
\noindent 
Measurements of the heavy quarkonium production in hadron collisions can be used to test predictions of quantum chromodynamics (QCD) in the
perturbative and non-perturbative regimes.
Several theoretical models have been developed within the framework of QCD to describe the quarkonium production
cross-section and polarisation as functions of the quarkonium transverse momentum, \pt, but none 
can simultaneously describe both of them~\cite{Brambilla}. 
Heavy quarkonia can be produced in three ways in $pp$ collisions: directly in the hard scattering, through feed-down
from higher-mass quarkonia
states, or via the decay of $b$ hadrons, with the first two of these being referred to as prompt production. In the
case of \psitwos mesons, the contribution from feed-down is negligible, allowing a straightforward comparison between
measurements of prompt production and predictions for direct contributions.

\par
The \psitwos meson has spin, parity and charge-parity quantum numbers, $J^{\mathrm{PC}}=1^{--}$ and its polarisation can be determined
by studying the angular distribution of muons in the $\decay{\psitwos}{\mup\mun}$ decays~\cite{Gottfried:1964nx,Faccioli:2010kd}. The distribution is described
by
\begin{equation}
  \frac{d^2N}{d\cos\theta\;d\phi}(\lambda_\theta,\lambda_{\theta\phi},\lambda_\phi)
  \propto 1+\lambda_\theta\cos^2\!\theta
  + \lambda_{\theta \phi}\sin2\theta\cos\phi
  + \lambda_\phi\sin^2\!\theta\cos2\phi,
  \label{theory1}
\end{equation}
where $\theta$ and $\phi$ are the polar and azimuthal angles of the $\mup$ direction 
in the rest frame of the \psitwos meson, respectively, 
and $\lambda_\theta,\lambda_{\theta\phi}$ and $\lambda_\phi$ are
the polarisation parameters to be determined from the data. 
The case of $(\lambda_{\theta}, \lambda_{\theta \phi}, \lambda_{\phi} ) = (1,\,0,\,0)$ or $(-1\,,0,\,0)$
corresponds to full transverse or longitudinal polarisation, respectively, 
while $(\lambda_\theta,\lambda_{\theta\phi},\lambda_\phi)=(0,\,0,\,0)$ 
corresponds to the unpolarised state.\footnote{
For a \psitwos meson in a pure spin state the three polarisation parameters cannot vanish simultaneously.
}
In this study of the \psitwos polarisation, two choices of polarisation frame are used.
These have a common definition of the $Y$-axis, taken to be the normal to the production plane, 
which is formed by the momentum of the \psitwos meson and the beam axis in the
rest frame of the colliding \lhc protons. 
The helicity frame~\cite{Jacob} uses the \psitwos momentum as the $Z$-axis. 
In the Collins-Soper frame~\cite{CollinsSoper} the $Z$-axis is chosen to be the bisector of the
angle between the two incoming proton beams in the rest frame of the \psitwos meson.  
In both frames, the $X$-axis is defined to complete a right-handed Cartesian coordinate system.
The commonly used frame-invariant variable $\lambda_\mathrm{inv}$ (see \cite{Faccioli1, Faccioli2})
is defined as 
\begin{equation}
\lambda_{\mathrm{inv}} = \frac{\lambda_\theta+3\lambda_\phi}{1-\lambda_\phi}.
\end{equation}
\indent

Two classes of theoretical models are compared with the measurements in this paper: 
the colour-singlet model (CSM)~\cite{Lansberg:2011hi} 
and non-relativistic QCD (NRQCD)~\cite{DetailedNRQCD:Bodwin,*PhysRevD.55.5853,Beneke:1996yw,COMJpsiPsi:Cacciari,COMPsi:Cho1,COMPsip:Braaten}, at next-to-leading order (NLO).
In the high-\pt region, where the quarkonium transverse momentum is much larger than its mass (in natural units), 
the CSM underestimates significantly the measured prompt \jpsi and \psitwos production cross-sections~\cite{Abe:1992ww, Aaij:2011jh, Aaij:2012ag}, while the
NRQCD model provides a good description of the \pt-dependent \jpsi and \psitwos cross-sections measured by
\lhcb~\cite{Aaij:2011jh, Aaij:2012ag} and \cms~\cite{Chatrchyan:2011kc}.
The CSM predicts large longitudinal polarisation for \jpsi and \psitwos mesons. 
On the other hand, in the NRQCD model, where quarkonium production is dominated by 
the gluon fragmentation process in the high-\pt region, 
the gluon is almost on-shell, leading to predictions of large transverse polarisations~\cite{Beneke:1996yw}. 
Precise measurements of the \jpsi polarisation at both the Tevatron~\cite{Abulencia:2007us} and the LHC~\cite{Abelev:2011md,Chatrchyan:2013cla,Aaij:2013nlm}, 
which show no significant longitudinal or transverse polarisations, 
do not support either the CSM or NRQCD predictions. 
\par
The prompt \psitwos polarisation has been measured by the \cdf
experiment~\cite{Abulencia:2007us} in $p\overline{p}$ collisions at $\sqrt{s}=1.96\tev$, 
and by the \cms experiment~\cite{Chatrchyan:2013cla} in $pp$ collisions at $\sqrt{s}=7\tev$, 
using the \decay{\psitwos}{\mup\mun} decay. 
The \cdf (\cms) measurement used \psitwos mesons in the kinematic range \mbox{$5<\pt<30\gevc$}
(\mbox{$14<\pt<50\gevc$}) and rapidity \mbox{$|y|<0.6$} ($|y|<1.5$).
The \cdf result for \mbox{$\pt>10\gevc$} is in strong disagreement
with the NRQCD prediction of large transverse polarisation.
At \cms, no evidence of large transverse or longitudinal \psitwos polarisation has been observed.
\par
This paper presents the measurement of the prompt \psitwos polarisation in $pp$ collisions 
at $\sqs=7\tev$, using data corresponding to an integrated luminosity of 1\invfb,
from \mbox{$\decay{\psitwos}{\mup\mun}$} decays.
The \psitwos polarisation parameters are determined using unbinned maximum likelihood fits 
to the two-dimensional angular distribution of the $\mup$ in the helicity and Collins-Soper frames. 
The measurement is performed in the \psitwos kinematic range \mbox{$3.5<\pt<15\gevc$} and $2.0<y<4.5$.

\section{LHCb detector and data sample} 
\label{detsample}
The \lhcb detector~\cite{Alves:2008zz} is a single-arm forward
spectrometer covering the \mbox{pseudorapidity} range \mbox{$2<\eta <5$},
designed for the study of particles containing \bquark or \cquark
quarks. The detector includes a high-precision tracking system
consisting of a silicon-strip vertex detector surrounding the $pp$
interaction region, a large-area silicon-strip detector located
upstream of a dipole magnet with a bending power of about
$4{\rm\,Tm}$, and three stations of silicon-strip detectors and straw
drift tubes placed downstream.
The combined tracking system provides a momentum measurement with
relative uncertainty that varies from 0.4\% at 5\gevc to 0.6\% at 100\gevc,
and impact parameter resolution of 20\mum for
tracks with large transverse momentum. Different types of charged hadrons are distinguished by information
from two ring-imaging Cherenkov detectors~\cite{LHCb-DP-2012-003}. Photon, electron and
hadron candidates are identified by a calorimeter system consisting of
scintillating-pad and preshower detectors, an electromagnetic
calorimeter and a hadronic calorimeter. Muons are identified by a
system composed of alternating layers of iron and multiwire
proportional chambers\cite{LHCb-DP-2012-002}.
\par
The trigger~\cite{LHCb-DP-2012-004} consists of a
hardware stage, based on information from the calorimeter and muon
systems, followed by a software stage, which applies full event reconstruction.
The hardware trigger requires the \pt of one muon candidate to be larger than \mbox{1.48\gevc}, 
or the product of the transverse momenta of two muon candidates to be larger than
\mbox{$1.68\gevct$}.
In a first stage of the software trigger, two oppositely charged muon candidates with \mbox{$\pt>0.5\gevc$} and momentum
\mbox{$p>6\gevc$} are selected and their invariant mass is required to be greater than $2.7\gevcc$. 
In a second stage of the software trigger, two muon candidates consistent with originating from a \psitwos decay are chosen
and additional criteria are applied to refine the sample of the \psitwos candidates as follows. 
The invariant mass of the candidate is required to be consistent with the known \psitwos mass~\cite{PDG2012}, and for 0.7\invfb
of data, the \pt of the candidate is required to be greater than 3.5\gevc.
\par
In the simulation, $pp$ collisions are generated using \pythia~\cite{Sjostrand:2006za} with a specific \lhcb
configuration~\cite{LHCb-PROC-2010-056}.  Decays of hadronic particles
are described by \evtgen~\cite{Lange:2001uf}, in which final state
radiation is generated using \photos~\cite{Golonka:2005pn}. The
interaction of the generated particles with the detector and its
response are implemented using the \geant
toolkit~\cite{Allison:2006ve, *Agostinelli:2002hh} as described in
Ref.~\cite{LHCb-PROC-2011-006}.  
The prompt charmonium production is simulated in \pythia according to the
leading order colour-singlet and colour-octet mechanisms \cite{LHCb-PROC-2010-056,Bargiotti:2007zz}, and the charmonium is generated without polarisation.

\section{Event selection}
\label{sec_sigsel}
The \psitwos candidates are reconstructed from pairs of good quality, oppositely charged particles that originate from a
common vertex. The \chisq probability of the vertex fit must be larger than 0.5\%. 
The transverse momentum of each particle is required to be greater than $1\gevc$.  
Both tracks must also be consistent with the muon hypothesis.
As in the measurement of $\jpsi$ polarisation~\cite{Aaij:2013nlm},
the significance $S_\tau$, which is defined as the reconstructed pseudo-decay time 
$\tau$ divided by its uncertainty, 
is used to distinguish between prompt \psitwos mesons and those from $b$-hadron decays.
The pseudo-decay time $\tau$ is defined as
\begin{equation}
\label{eq:tau}
 \tau\equiv\frac{(z_{\psitwos}-z_{\mathrm{PV}})\cdot{}M_{\psitwos}}{p_z},
\end{equation}
where $z_{\psitwos}$ ($z_{\mathrm{PV}})$ is the position of 
the $\psitwos$ decay vertex (the associated primary vertex) in the $z$-direction, 
$M_{\psitwos}$ is the known $\psitwos$ mass, and $p_z$ is the measured $z$ component 
of the $\psitwos$ momentum in the centre-of-mass frame of the $pp$ collision. The $z$-axis of the \lhcb coordinate
system is defined as the beam direction in the \lhcb detector region.
The \psitwos mesons from $b$-hadron decays tend to have large values of $S_\tau$.
The requirement $S_\tau<4$ reduces the fraction of the selected non-prompt \psitwos mesons from about 20\% to 3\%. 
\begin{figure}[!t]
\centering
\includegraphics[width=0.7\textwidth]{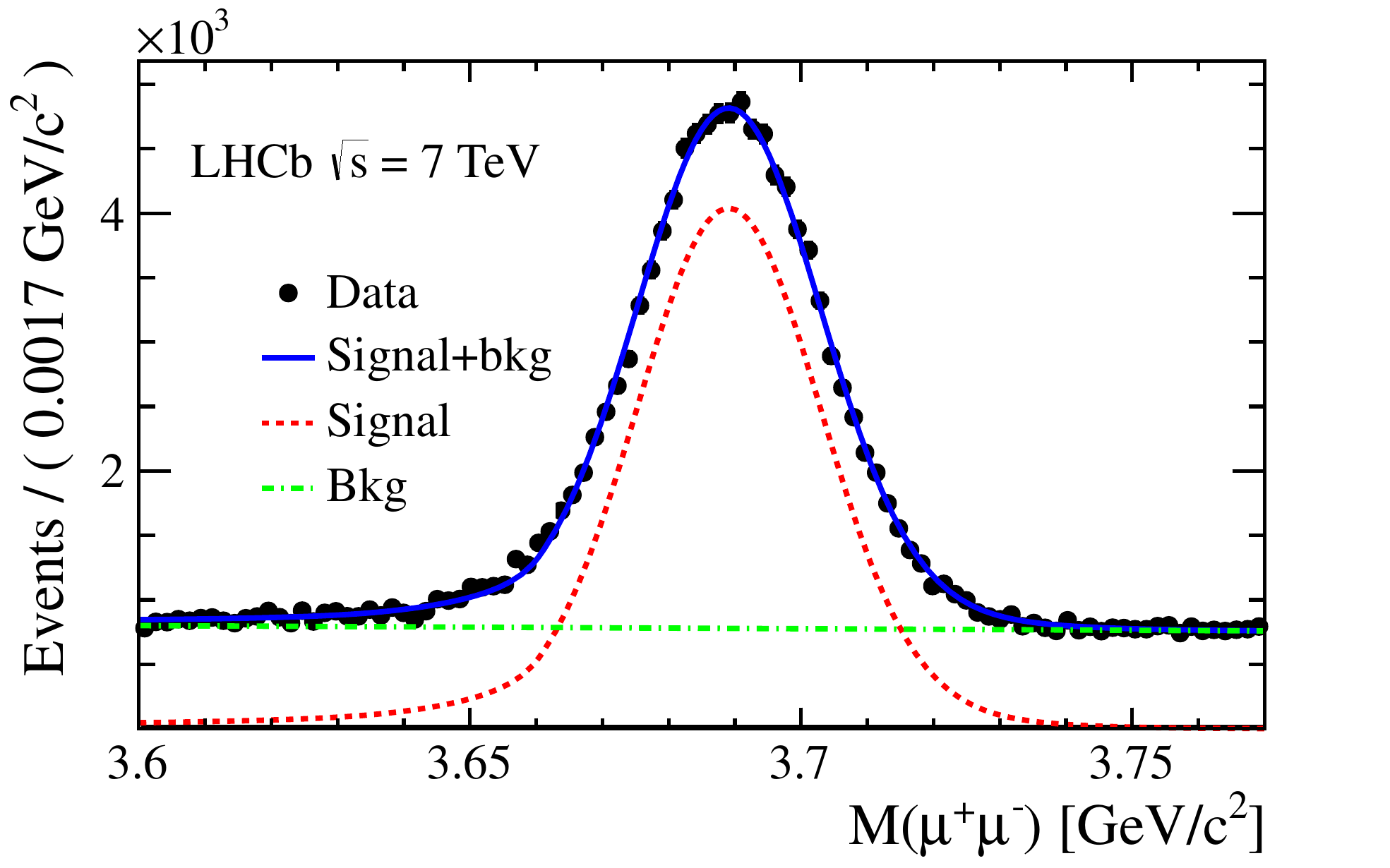}
\caption{
   \small{
      Invariant mass distribution of \psitwos candidates in the kinematic region \mbox{$5<\pt<7\gevc$} 
      and \mbox{$2.5<y<3.0$}. 
      The solid blue line is the total fit function, the  dot-dashed green line
      represents the linear background function and the red dashed line 
      is the combination of the two CB functions.} 
   }
   \label{fig:mass}
\end{figure}
\par
The analysis is performed in five \pt and five $y$ bins of the \psitwos meson.
As an example, the invariant mass distribution of \psitwos candidates for 
\mbox{$5<\pt<7\gevc$} and \mbox{$3.0<y<3.5$} is shown in Fig.~\ref{fig:mass}.
In each kinematic bin, the mass distribution is fitted with a combination of two Crystal Ball (CB) functions~\cite{Skwarnicki:1986xj} with a
common peak position for the signal and a linear function for the combinatorial background.
The relative fractions of the narrower and broader CB functions are fixed to 0.9 and 0.1, respectively, determined from
simulation.
\par
Using the results of the fit to the mass distribution, the \sWeight $w_i$ for each candidate $i$ to be signal
is computed by means of the \sPlot technique~\cite{Pivk:2004ty}. 
The correlation between the invariant mass of the \psitwos candidates and the muon angular variables is found to be negligible, 
and the {\em sWeights} are used to subtract the background from the angular distribution. 

\section{Polarisation fit}
\label{polfit}
\begin{figure}[!t]
\centering
\includegraphics[width=0.7\textwidth]{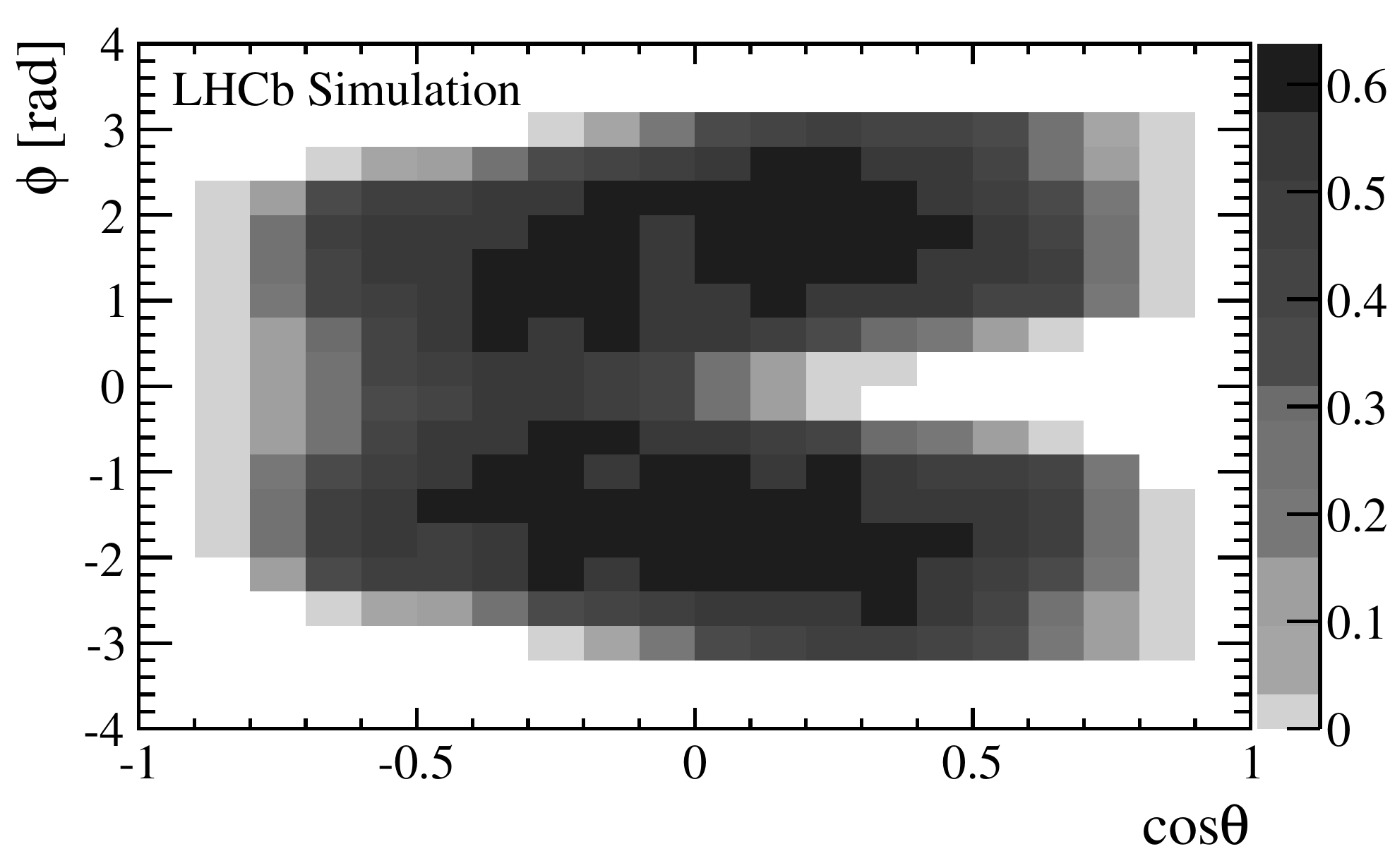}
\caption{\small{Detection efficiency in arbitrary units as a function of $\cos \theta$ and  $\phi$ in the helicity frame
      for  \psitwos mesons in the range \mbox{$5<\pt<7\gevc$} and $2.5<y<3.0$.
}
   \label{fig:acceptance}}
\end{figure}

The polarisation parameters are determined from a fit to the ($\cos\theta,\phi$) angular
distribution of the \decay{\psitwos}{\mup\mun}~signal candidates in each kinematic bin of the \psitwos meson
independently.
The angular distribution described by Eq.~\ref{theory1} is modified by the detection efficiency $\epsilon$,
which varies as a function of the angular variables ($\cos\theta,\phi$).
In each kinematic bin, $\epsilon$ is obtained from
a sample of simulated unpolarised \decay{\psitwos}{\mup\mun} decays, where $\cos\theta$ and $\phi$ are generated
according to uniform distributions. As an example, Fig.~\ref{fig:acceptance} shows the efficiency in the helicity frame
for \psitwos candidates in the kinematic bin $5<\pt<7\gevc$ and $2.5<y<3.0$. For smaller (larger) \pt and $y$ values, the
coverage of the reconstructed muon angular variables is narrower (broader).
In the regions $\vert\cos\theta\vert \approx 1$, and $\vert\phi\vert \approx 0$ or $\pi$, the efficiency is lower
because one of the two muons is likely to escape the LHCb detector acceptance.

Combining the angular distribution given in Eq.~\ref{theory1} with the efficiency, the logarithm of the likelihood
function~\cite{Xie:2009ar}, in each \pt and $y$ bin, is defined as

\vspace{-1.0em}
\begin{eqnarray}
\ln L&=&\alpha\sum^{N_{\mathrm{tot}}}_{i=1}w_i\times
         \ln\left[\frac{P(\cos\theta_{i},\phi_{i}\vert\lambda_{\theta},
        \lambda_{\theta\phi}, \lambda_{\phi})\;\epsilon(\cos\theta_{i},\phi_{i})}{N(\lambda_{\theta},
        \lambda_{\theta\phi}, \lambda_{\phi})}\right], \label{likelihood2}
\end{eqnarray}
where $P(\cos\theta_{i},\phi_{i}\vert\lambda_{\theta},\lambda_{\theta\phi}, \lambda_{\phi}) 
         \equiv 1+\lambda_\theta \cos^2 \theta_{i}
               + \lambda_{\theta \phi}\sin 2\theta_{i} \cos \phi_{i}
               + \lambda_\phi\sin^2 \theta_{i} \cos 2\phi_{i}$,
$w_i$ is the \sWeight, and $N_\mathrm{tot}$ is the number of $\psitwos$ candidates in the data. 
The global factor \mbox{$\alpha\equiv\sum_{i=1}^{N_\mathrm{tot}}w_i/\sum_{i=1}^{N_\mathrm{tot}}w_i^2$}
is introduced to estimate correctly the statistical uncertainty for the weighted likelihood function.
The normalisation $N(\lambda_{\theta}, \lambda_{\theta\phi}, \lambda_{\phi})$ is defined as

\vspace{-1.0em}
\begin{eqnarray}\label{Normalization1}
N(\lambda_{\theta},\lambda_{\theta\phi},\lambda_{\phi}) &=& \int d\Omega 
P(\cos\theta,\phi\vert\lambda_{\theta}, \lambda_{\theta\phi}, \lambda_{\phi})\times\epsilon(\cos\theta,\phi) \nonumber\\
      &=&C\sum_{j=1}^{M_\mathrm{tot}}
P(\cos\theta_j,\phi_j\vert\lambda_{\theta}, \lambda_{\theta\phi}, \lambda_{\phi}),
\end{eqnarray}
where the sum extends over the $M_\mathrm{tot}$ candidates in the simulated sample and $C$ is a constant factor.
The last equality holds because the $(\cos\theta,\phi)$ two-dimensional distribution for the fully simulated 
unpolarised \psitwos mesons is the same as the efficiency
$\epsilon(\cos\theta,\phi)$ up to a constant global factor.
\par
\begin{figure}[!t]
\begin{minipage}[t]{0.5\textwidth}
\centering
\includegraphics[width=1.0\textwidth]{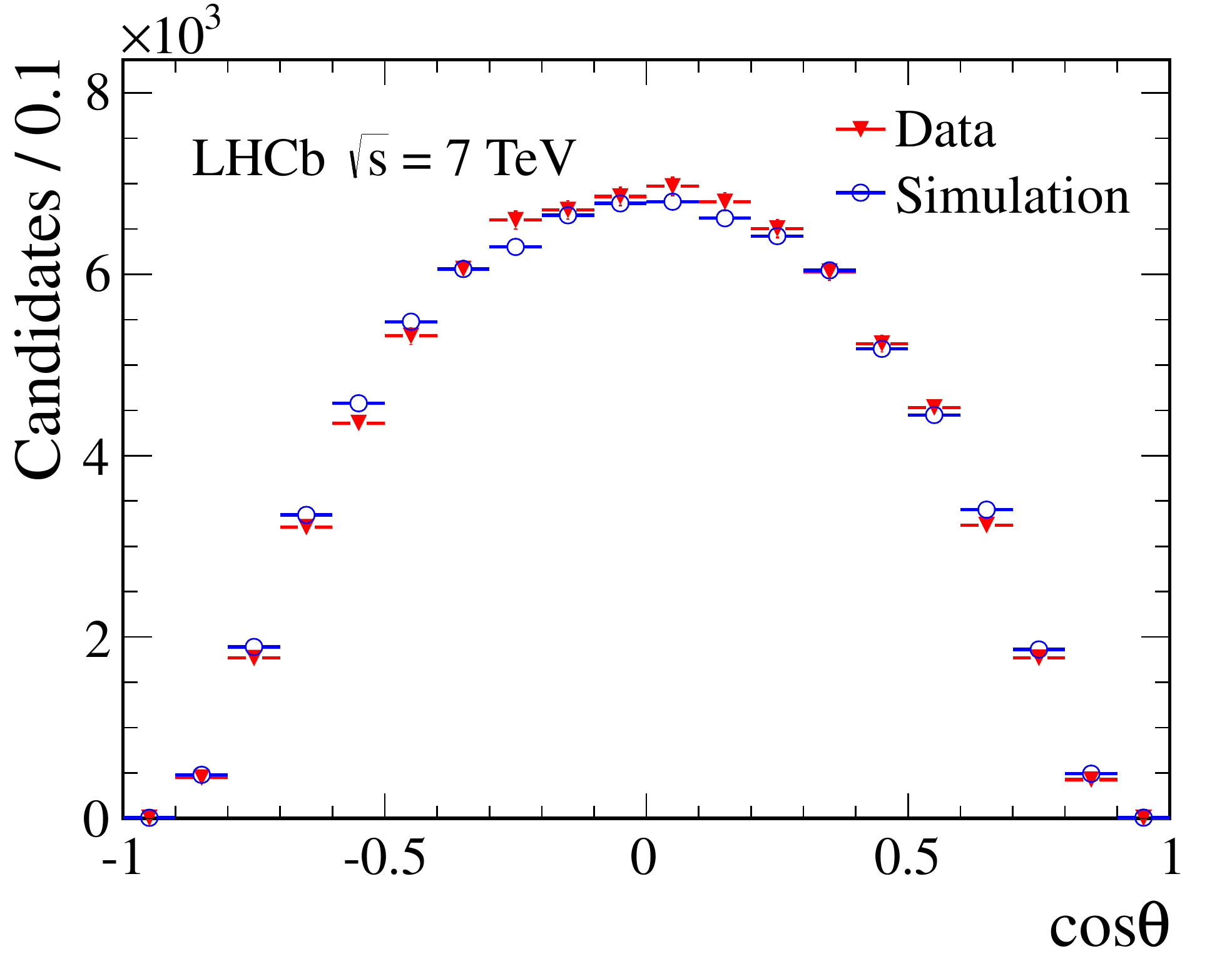}
\end{minipage}
\begin{minipage}[t]{0.5\textwidth}
\centering
\includegraphics[width=1.0\textwidth]{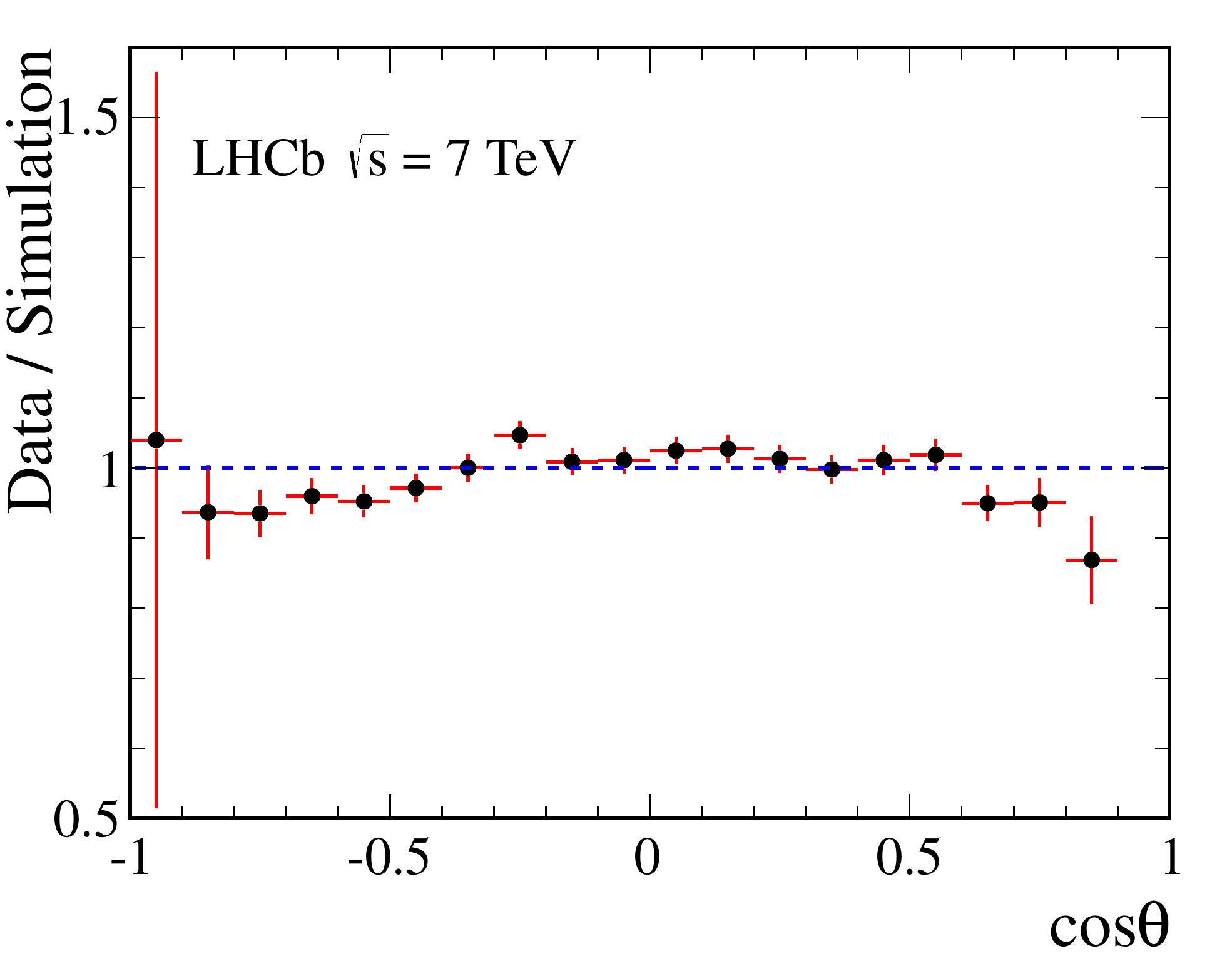}
\end{minipage}
\caption{\small{(\textit{Left}) Distributions of $\cos\theta$ in the helicity frame for \jpsi mesons from
   $\Bp \rightarrow \jpsi K^+$ decays in data (filled triangles) and in the simulated
      sample (open circles) and (\textit{right}) their ratio after the weighting based on the $\Bp$ and \jpsi kinematic properties.}
   \label{fig:B2JpsiKMAngleB}}
   \end{figure}

The angular efficiency is validated in data by using muons from \mbox{\decay{\Bp}{\jpsi\Kp}} decays.  Due to angular momentum conservation, 
the \jpsi meson produced in this channel is longitudinally polarised in the $\Bp$
meson rest frame. After reweighting the kinematic properties of the simulated \Bp and \jpsi mesons to reproduce the data, 
the remaining differences of the angular distributions between the reweighted simulation sample and the data 
are attributed to imperfections in the modelling of the detector response.
Figure~\ref{fig:B2JpsiKMAngleB} compares the $\cos\theta$ distributions in data for \decay{\Bp}{\jpsi\Kp} candidates in the
helicity frame with simulated data after reweighting.  
The efficiency for simulated events is overestimated for \jpsi candidates with $|\cos\theta|>0.5$, therefore it 
is corrected as a function of $p_\mu$ and $y_\mu$, the momentum and the rapidity of the
muon in the centre-of-mass frame of $pp$ collisions. 
The normalisation of Eq.~\ref{Normalization1} is calculated by
assigning a weight to each candidate as the product of the
weights for \mup and \mun depending on their $p_\mu$ and $y_\mu$ values.

\section{Systematic uncertainties}
\label{sec:Systematics}
\noindent
Sources of systematic uncertainty are considered for each of the four observables $\lla$, $\llb$,
$\llc$ and $\lli$ in both the Collins-Soper and helicity frames.  
In the Collins-Soper frame, the
overall systematic uncertainties are found to be comparable for each of these observables in most kinematic bins, while for
the helicity frame the systematic uncertainties assigned to $\llb$ and $\llc$ are typically a
factor of 2--3 smaller than those estimated for $\lla$ and $\lli$.
For each of the main sources of systematic uncertainty, Table~\ref{tab:systematic} shows
the range of values assigned over all kinematic bins, and their average. 
The total systematic uncertainties for each of the four observables can be found in Tables~\ref{tab:PolarisationResultHX}
and~\ref{tab:PolarisationResultCS}.
\par
The dominant systematic uncertainty is due to the size of the \decay{\Bp}{\jpsi\Kp} control
sample. This leads to non-negligible statistical uncertainties in the correction factors that are applied to simulated
events in bins of $p_\mu$ and $y_\mu$. 
The uncertainty on a given correction factor is estimated by varying it by one standard deviation of its statistical
uncertainty, while keeping all other factors at their central values.
The polarisation parameters are recalculated and the change relative to their default values is considered as the
contribution from this factor to the systematic uncertainty.  This procedure is repeated for all bins of $p_\mu$ and $y_\mu$,
and the sum in quadrature of all these independent contributions is taken as the total systematic uncertainty.
\par
The limited size of the sample of simulated events introduces an uncertainty on the normalisation $N(\lambda_\theta,
\lambda_{\theta\phi},\lambda_\phi)$, and this uncertainty is propagated to the polarisation parameters. 
\par
The uncertainty of the~\sWeight~of each candidate used for the background subtraction is a source of uncertainty on the polarisation
parameters. The effect is studied by comparing the default polarisation parameters with those obtained when varying the
definition of the models used to fit the mass distributions and re-evaluating the \sWeight for each candidate.
Several alternative fitting models are studied, including an exponential function for the background mass distribution, only one CB
function for the signal mass distribution, or shapes for signal and background mass distributions fixed to those obtained from fits 
to the mass distributions in sub-regions of the $(\cos\theta,\phi)$ distribution space.
The largest variation with respect to the default result is assigned as the systematic uncertainty. 
\par
In each kinematic bin, discrepancies between data and simulation in the \psitwos \pt and $y$ distributions
introduce an additional uncertainty.  This is evaluated by comparing the default polarisation results with
those determined after the \psitwos kinematic distribution in the simulation is weighted to that in data.
The difference between the two results is quoted as a systematic uncertainty contribution. 
\par
The uncertainty due to the contamination of \psitwos candidates from $b$-hadron decays ($3\%$) is
determined by relaxing the $S_\tau$ selection and studying the variations of the polarisation parameters. 
\par
With the exception of the effects due to the differences in the \psitwos kinematic spectrum and the 
size of the sample of simulated events, correlations are expected among \psitwos kinematic bins. 
The correlation between these systematic uncertainties in adjacent bins could be as large as
50\%, as the final state muons may have similar momentum and rapidity.
For each kinematic bin, the total systematic uncertainty is calculated as the quadratic sum of the various sources of systematic uncertainties assuming no correlation within each kinematic bin. 

\begin{table}
  \caption{\small
    Sources of systematic uncertainties on the polarisation 
    parameter $\lambda_\theta$ in the helicity and Collins-Soper frames.
    For each type of uncertainty, the average and the range over  all \psitwos kinematic bins are shown.}
  \label{tab:systematic}
  \begin{center}
    \begin{tabular}{lcc}
        \toprule[1.pt]
      \multirow{2}{*}{Source} & Helicity frame &  Collins-Soper frame \\
      & Average (range) &  Average (range) \\
      \midrule[1.pt]
      Efficiency correction        & 0.055 (0.034 -- 0.126) & 0.035 (0.019 -- 0.078) \\
      Simulation sample size       & 0.034 (0.015 -- 0.103) & 0.023 (0.010 -- 0.094) \\
      Fit to mass distribution     & 0.008 (0.001 -- 0.134) & 0.007 (0.001 -- 0.188) \\
      \psitwos kinematic modelling & 0.018 (0.000 -- 0.085) & 0.016 (0.000 -- 0.074) \\
      \bquark-hadron contamination & 0.014 (0.002 -- 0.035) & 0.013 (0.002 -- 0.063) \\
      \bottomrule[1.pt]
    \end{tabular}
  \end{center}
\end{table}

\section{Results}
\label{results}
\begin{figure}[!t]
  \begin{minipage}[t]{0.5\textwidth}
    \centering
    \includegraphics[width=1.0\textwidth]{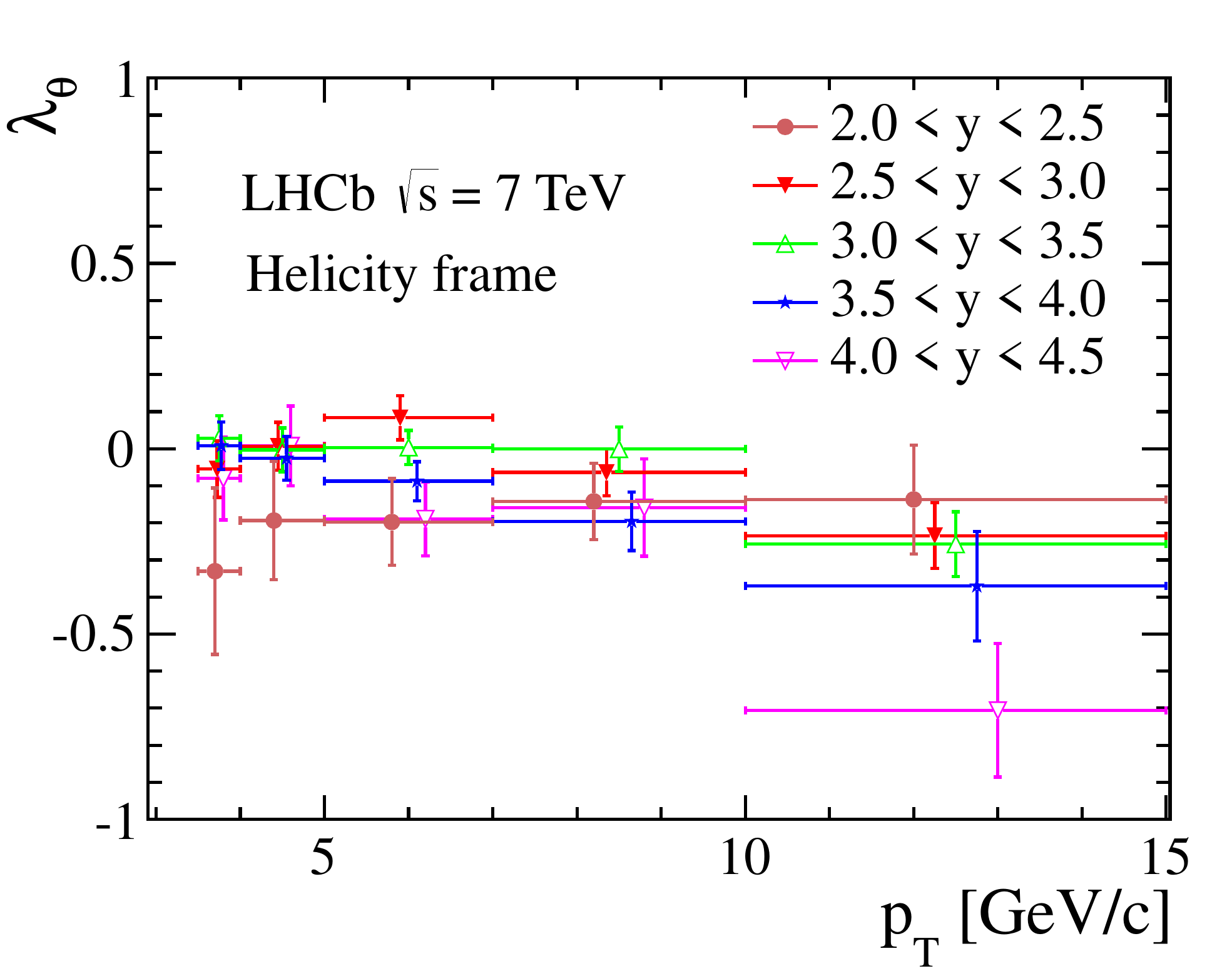}
  \end{minipage}
  \begin{minipage}[t]{0.5\textwidth}
    \centering
    \includegraphics[width=1.0\textwidth]{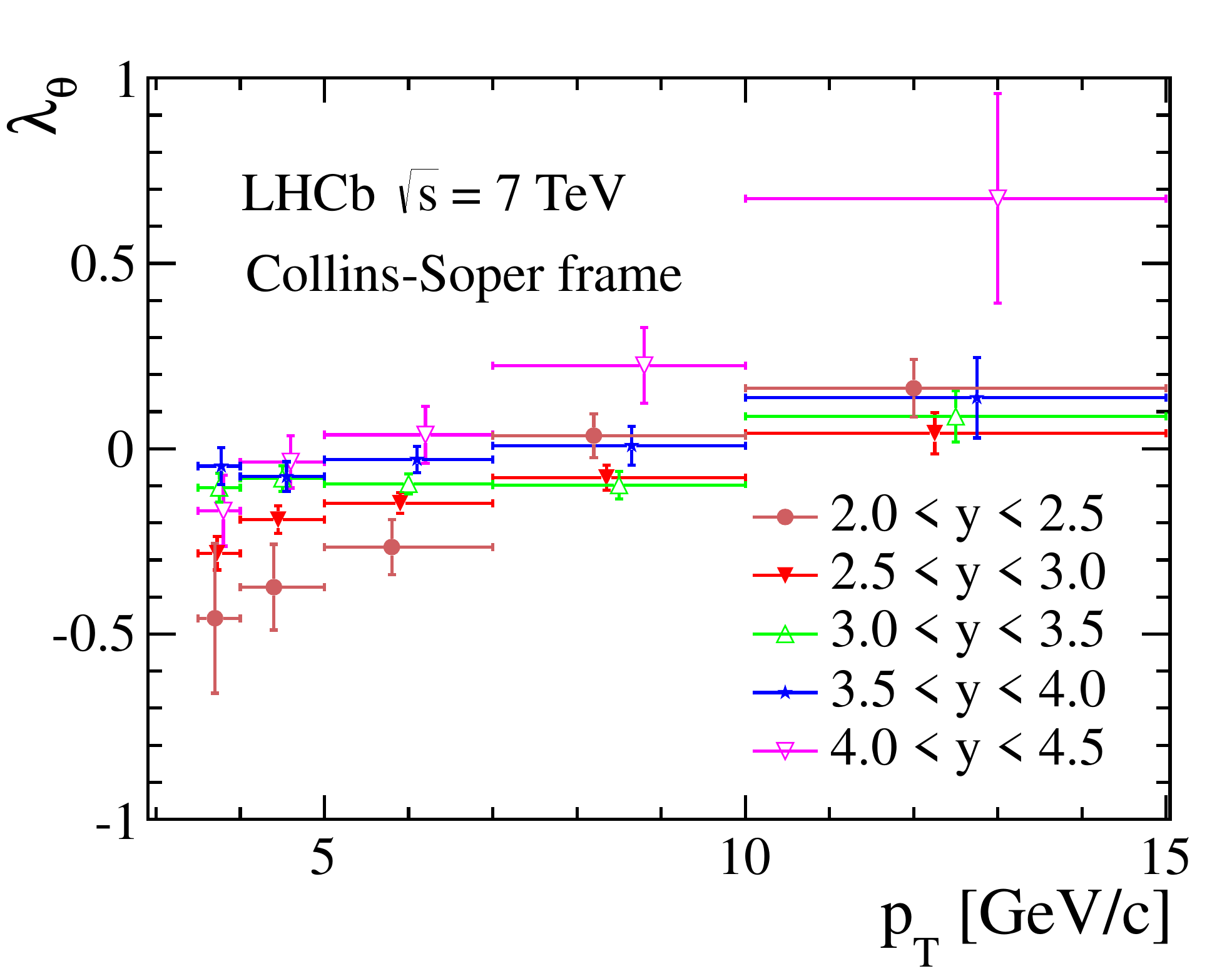}
  \end{minipage}
  \begin{minipage}[t]{1.0\textwidth}
    \centering
    \includegraphics[width=0.5\textwidth]{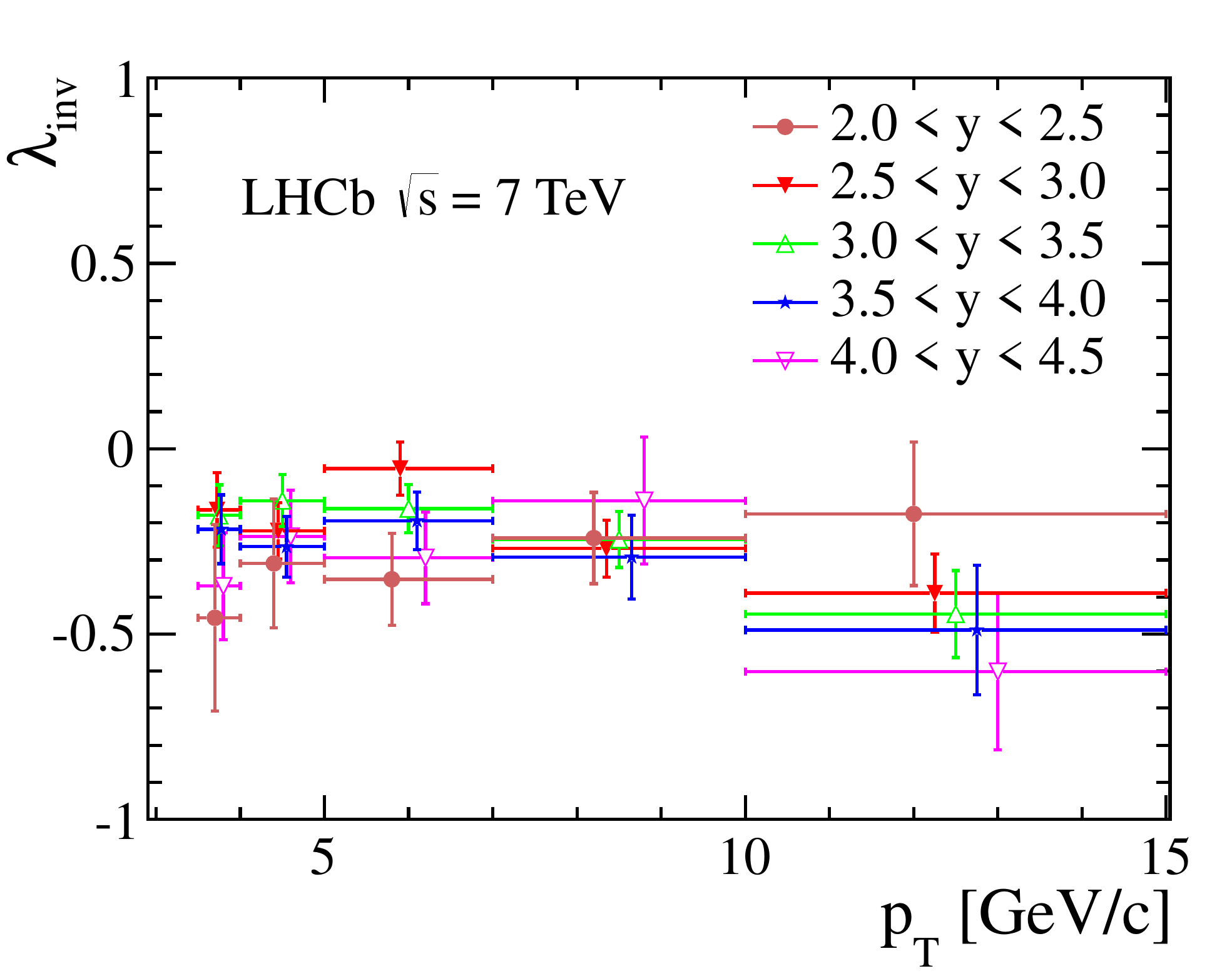}
  \end{minipage}
  \caption{\small
  Polarisation parameters for prompt \psitwos mesons as a function of \pt, in
  five rapidity intervals, (\textit{top left}) $\lla$ and (\textit{bottom}) $\lli$,
  measured in the helicity frame, and (\textit{top right}) $\lla$ in the
  Collins-Soper frame.
    The uncertainties on data points are the sum in quadrature of statistical and systematic uncertainties.
    The horizontal bars represent the width of the \pt bins for the \psitwos meson.
    The data points for each rapidity interval are displaced horizontally to improve visibility.
    }
  \label{LambdaTheta}
\end{figure}

The results for the polarisation parameters \lla, \llb, \llc and \lli, and their uncertainties, in each \pt and $y$ bin
of the prompt \psitwos meson sample, are
reported in Tables~\ref{tab:PolarisationResultHX} and~\ref{tab:PolarisationResultCS} for the helicity and the
Collins-Soper frames, respectively.   
The systematic uncertainties are similar in size to the statistical uncertainties.
The parameters \lla and \lli are also shown in Fig.~\ref{LambdaTheta} 
as functions of the \pt of the \psitwos mesons, for different $y$ bins.
\par
The frame-invariant polarisation parameter \lli is consistent with a negative polarisation with no strong 
dependence on the \pt and $y$ of the \psitwos meson.
The values and uncertainties of \lli that are measured in the helicity and the Collins-Soper frames are in good agreement with each
other, with differences much smaller than the statistical uncertainties.  
In the Collins-Soper frame, \lla takes small negative values especially in the low-\pt region and increases with
\pt. This trend is more significant for the extreme $y$ bins.
In the helicity frame, the polarisation parameter \lla is consistent with zero, with no significant dependence on \pt or $y$ of the \psitwos
meson.
The polarisation parameters \llb and \llc are consistent with zero in both the helicity and Collins-Soper
frames, and their absolute values are below 0.1 for most of the kinematic bins.
\par
\begin{figure}[!t]
  \begin{center}
   \includegraphics[angle=0,width=0.8\textwidth]{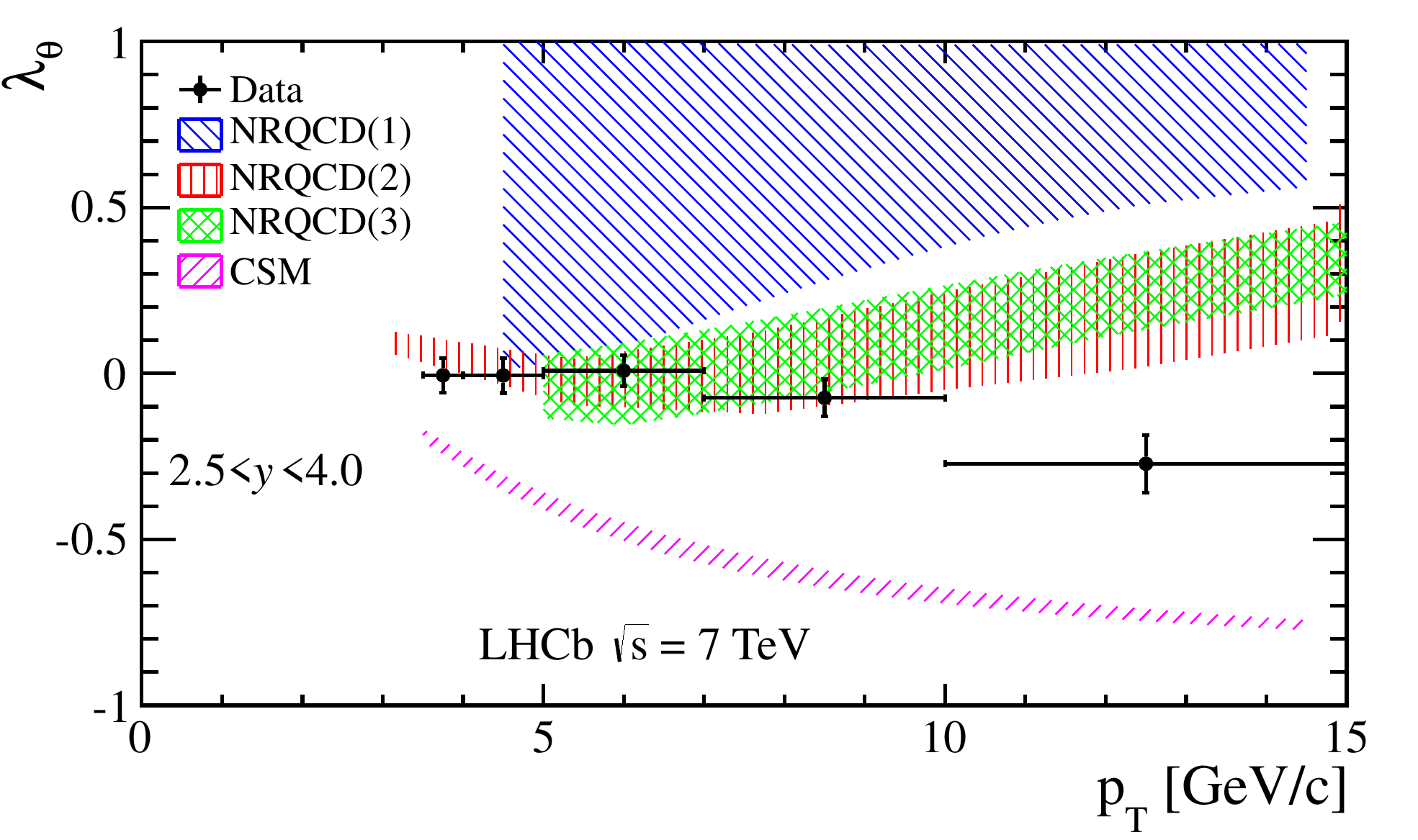}
   \caption{\small
   Polarisation parameter \lla of the prompt \psitwos meson
   in the helicity frame as a function of \pt, in the rapidity
   range $2.5<y<4$.  
   The predictions of NLO CSM~\cite{Butenschoen:2012px} and three NLO NRQCD
   models (1)~\cite{Butenschoen:2012px}, (2)~\cite{Gong:2012ug} and (3)~\cite{Chao:2012iv} are also shown.
   Uncertainties on data are the sum in quadrature of the statistical
   and systematic uncertainties.  The horizontal bars represent
   the width of \pt bins for the \psitwos meson.
     }
   \label{CompareWithTheory}
  \end{center}
\end{figure}

In Fig.~\ref{CompareWithTheory}, the measured values of $\lambda_\theta$ in the helicity frame as a function of \pt of the
\psitwos meson, integrating over the rapidity range $2.5<y<4.0$, are
compared with the predictions of the CSM~\cite{Butenschoen:2012px} and
NRQCD~\cite{Butenschoen:2012px,Gong:2012ug,Chao:2012iv} models at NLO.
Our results disfavour the CSM calculations, in which the \psitwos meson is significantly longitudinally polarised.
The three NRQCD calculations in Refs.~\cite{Butenschoen:2012px,Gong:2012ug,Chao:2012iv}, which 
use different selections of experimental data to determine the non-perturbative matrix elements, provide a good
description of our measurements in the low-\pt region.
However, the prediction of increasing polarisation with \pt in these models is not supported by the \lhcb data.

\section{Conclusion} \label{conclusion}
The polarisation of prompt \psitwos mesons is measured as a function of the \psitwos \pt and $y$ in the range
$3.5<\pt<15\gevc$ and $2.0<y<4.5$, in $pp$ collisions at $\sqrt{s}= 7\tev$. 
The analysis is performed using data corresponding to an integrated luminosity of 1.0~\invfb, collected by the LHCb
experiment in 2011.
The polarisation parameters \lla, \llb, \llc and \lli are determined in the helicity and Collins-Soper frames by studying the angular distribution of the
two muons produced in the \decay{\psitwos}{\mup \mun} decay.
\par
The frame-independent observable \lli is consistent with a negative polarisation. The measured values of \llb and \llc are small over the
accessible kinematic range. 
The \lla distribution in the helicity frame shows that the \psitwos meson
exhibits neither large transverse nor longitudinal polarisation.
Although a direct comparison with previous
measurements by \cms and \cdf is not possible due to the different
kinematic ranges, all results disfavour large polarisation in the high-\pt region.
The prompt \psitwos polarisation measured at \lhcb disagrees with the CSM 
predictions both in the size of the polarisation parameters and
the $\pt$ dependence.  While the NRQCD models provide a good description of the \lhcb data in
the low-\pt region, the predicted transverse polarisation at high-\pt is not observed.

\section*{Acknowledgements}
 
\noindent 
We wish to thank M. Butensch{\"o}n, B. Gong, H.-S. Shao and Y.-Q. Ma 
for providing us with the theoretical calculations and helpful discussions.
We express our gratitude to our colleagues in the CERN
accelerator departments for the excellent performance of the LHC. We
thank the technical and administrative staff at the LHCb
institutes. We acknowledge support from CERN and from the national
agencies: CAPES, CNPq, FAPERJ and FINEP (Brazil); NSFC (China);
CNRS/IN2P3 and Region Auvergne (France); BMBF, DFG, HGF and MPG
(Germany); SFI (Ireland); INFN (Italy); FOM and NWO (The Netherlands);
SCSR (Poland); MEN/IFA (Romania); MinES, Rosatom, RFBR and NRC
``Kurchatov Institute'' (Russia); MinECo, XuntaGal and GENCAT (Spain);
SNSF and SER (Switzerland); NAS Ukraine (Ukraine); STFC (United
Kingdom); NSF (USA). We also acknowledge the support received from EPLANET and the
ERC under FP7. The Tier1 computing centres are supported by IN2P3
(France), KIT and BMBF (Germany), INFN (Italy), NWO and SURF (The
Netherlands), PIC (Spain), GridPP (United Kingdom).
We are indebted to the communities behind the multiple open source software packages on which we depend.
We are also thankful for the computing resources and the access to software R\&D tools provided by Yandex LLC (Russia).

\clearpage



\begin{landscape}
  \begin{table}
    \caption{\small
      Measured prompt \psitwos polarisation parameters \lla, \llb, \llc and \lli in bins of \pt and $y$
      in the helicity frame.  The first uncertainty is statistical and the is second systematic.}
    \label{tab:PolarisationResultHX}
    \begin{center}
      \small
      \begin{tabular}{ccr@{$\pm$}c@{$\pm$}cr@{$\pm$}c@{$\pm$}cr@{$\pm$}c@{$\pm$}cr@{$\pm$}c@{$\pm$}cr@{$\pm$}c@{$\pm$}c}
        \toprule[1.0pt]
        \pt ($\mathrm{Ge\kern -0.1em V\!/}c$) & $\lambda$ & \multicolumn{3}{c}{$2.0<y<2.5$} & \multicolumn{3}{c}{$2.5<y<3.0$} & \multicolumn{3}{c}{$3.0<y<3.5$} & \multicolumn{3}{c}{$3.5<y<4.0$} & \multicolumn{3}{c}{$4.0<y<4.5$}  \\
        \midrule[1.0pt]
\multirow{4}{*}{3.5--4}&$\lambda_\theta$   &$-$0.331&0.174&0.142  &$-$0.055&0.052&0.056 &0.028&0.040&0.046  &0.008&0.040&0.050  &$-$0.080&0.063&0.092\\
&$\lambda_{\theta\phi}$ &$-$0.233&0.076&0.086  &$-$0.172&0.021&0.026  &$-$0.039&0.020&0.023  &0.007&0.021&0.028  &$-$0.048&0.036&0.049\\
&$\lambda_\phi$         &$-$0.049&0.036&0.037  &$-$0.039&0.017&0.024  &$-$0.074&0.018&0.022  &$-$0.081&0.022&0.027  &$-$0.110&0.043&0.047\\
&$\lambda_\mathrm{inv}$ &$-$0.456&0.195&0.160  &$-$0.165&0.063&0.078  &$-$0.180&0.054&0.063  &$-$0.217&0.057&0.073  &$-$0.371&0.089&0.114\\
\midrule[1.0pt]
\multirow{4}{*}{4--5}&$\lambda_\theta$   &$-$0.194&0.113&0.113  &0.007&0.038&0.052  &$-$0.003&0.028&0.052  &$-$0.026&0.029&0.052  &0.007&0.050&0.095\\
&$\lambda_{\theta\phi}$ &$-$0.238&0.049&0.053  &$-$0.086&0.016&0.023  &$-$0.026&0.015&0.021  &0.003&0.017&0.025  &0.023&0.027&0.043\\
&$\lambda_\phi$         &$-$0.043&0.023&0.024  &$-$0.082&0.012&0.014  &$-$0.048&0.012&0.023  &$-$0.087&0.016&0.025  &$-$0.088&0.033&0.035\\
&$\lambda_\mathrm{inv}$ &$-$0.309&0.126&0.120  &$-$0.222&0.045&0.060  &$-$0.140&0.042&0.057  &$-$0.265&0.044&0.070  &$-$0.237&0.072&0.102\\
\midrule[1.0pt]
\multirow{4}{*}{5--7}&$\lambda_\theta$   &$-$0.198&0.074&0.091  &0.083&0.030&0.051  &0.003&0.024&0.039  &$-$0.088&0.024&0.046  &$-$0.189&0.039&0.092\\
&$\lambda_{\theta\phi}$ &$-$0.164&0.031&0.039  &$-$0.072&0.013&0.018  &$-$0.026&0.013&0.020  &0.002&0.015&0.026  &0.044&0.025&0.051\\
&$\lambda_\phi$         &$-$0.058&0.014&0.021  &$-$0.046&0.009&0.013  &$-$0.058&0.009&0.018  &$-$0.038&0.012&0.019  &$-$0.039&0.025&0.028\\
&$\lambda_\mathrm{inv}$ &$-$0.352&0.080&0.094  &$-$0.054&0.039&0.060  &$-$0.162&0.035&0.054  &$-$0.195&0.040&0.067  &$-$0.294&0.065&0.105\\
\midrule[1.0pt]
\multirow{4}{*}{7--10}&$\lambda_\theta$   &$-$0.142&0.066&0.079  &$-$0.064&0.034&0.053  &$-$0.001&0.032&0.051  &$-$0.196&0.033&0.071  &$-$0.159&0.058&0.118\\
&$\lambda_{\theta\phi}$ &0.044&0.028&0.034  &0.002&0.014&0.021  &0.008&0.016&0.023  &0.003&0.019&0.031  &0.124&0.037&0.058\\
&$\lambda_\phi$         &$-$0.036&0.014&0.017  &$-$0.075&0.010&0.012  &$-$0.088&0.011&0.012  &$-$0.036&0.014&0.018  &0.007&0.030&0.031\\
&$\lambda_\mathrm{inv}$ &$-$0.241&0.079&0.095  &$-$0.269&0.043&0.064  &$-$0.245&0.043&0.062  &$-$0.292&0.052&0.100  &$-$0.140&0.101&0.138\\
\midrule[1.0pt]
\multirow{4}{*}{10--15}&$\lambda_\theta$   &$-$0.137&0.080&0.123  &$-$0.235&0.047&0.075  &$-$0.258&0.048&0.073  &$-$0.371&0.059&0.135  &$-$0.706&0.081&0.161\\
&$\lambda_{\theta\phi}$ &0.157&0.034&0.050  &0.045&0.020&0.026  &0.094&0.023&0.032  &0.052&0.031&0.054  &0.104&0.059&0.079\\
&$\lambda_\phi$         &$-$0.014&0.021&0.022  &$-$0.059&0.017&0.011  &$-$0.074&0.020&0.014  &$-$0.047&0.027&0.020  &0.044&0.053&0.048\\
&$\lambda_\mathrm{inv}$ &$-$0.176&0.103&0.164  &$-$0.390&0.062&0.086  &$-$0.446&0.067&0.096  &$-$0.489&0.089&0.150  &$-$0.601&0.162&0.136\\
        \bottomrule[1.0pt]

      \end{tabular}
    \end{center}
  \end{table}
\end{landscape}

\begin{landscape}
  \begin{table}
    \caption{\small
      Measured prompt \psitwos polarisation parameters \lla, \llb, \llc and \lli in bins of \pt and $y$
      in the Collins-Soper frame. The first uncertainty is statistical and the second is systematic.}
    \label{tab:PolarisationResultCS}
    \begin{center}
      \small
      \begin{tabular}{ccr@{$\pm$}c@{$\pm$}cr@{$\pm$}c@{$\pm$}cr@{$\pm$}c@{$\pm$}cr@{$\pm$}c@{$\pm$}cr@{$\pm$}c@{$\pm$}c}
        \toprule[1.0pt]
        \pt ($\mathrm{Ge\kern -0.1em V\!/}c$) & $\lambda$ & \multicolumn{3}{c}{$2.0<y<2.5$} & \multicolumn{3}{c}{$2.5<y<3.0$} & \multicolumn{3}{c}{$3.0<y<3.5$} & \multicolumn{3}{c}{$3.5<y<4.0$} & \multicolumn{3}{c}{$4.0<y<4.5$}  \\
        \midrule[1.pt]
\multirow{4}{*}{3.5--4}&$\lambda_\theta$   &$-$0.457&0.142&0.144  &$-$0.282&0.026&0.036  &$-$0.105&0.023&0.031  &$-$0.047&0.028&0.041  &$-$0.168&0.058&0.076\\
&$\lambda_{\theta\phi}$ &0.141&0.088&0.065  &0.018&0.027&0.031  &$-$0.043&0.022&0.027  &$-$0.038&0.024&0.032  &$-$0.010&0.044&0.059\\
&$\lambda_\phi$         &$-$0.003&0.039&0.028  &0.040&0.020&0.023  &$-$0.027&0.021&0.024  &$-$0.061&0.023&0.028  &$-$0.076&0.031&0.045\\
&$\lambda_\mathrm{inv}$ &$-$0.465&0.194&0.179  &$-$0.169&0.062&0.068  &$-$0.180&0.054&0.062  &$-$0.218&0.057&0.076  &$-$0.368&0.089&0.118\\
\midrule[1.0pt]
\multirow{4}{*}{4--5}&$\lambda_\theta$   &$-$0.374&0.077&0.086  &$-$0.192&0.019&0.032  &$-$0.080&0.017&0.030  &$-$0.075&0.020&0.035  &$-$0.035&0.042&0.056\\
&$\lambda_{\theta\phi}$ &0.103&0.059&0.062  &$-$0.020&0.019&0.028  &$-$0.010&0.015&0.035  &$-$0.027&0.017&0.032  &$-$0.047&0.034&0.057\\
&$\lambda_\phi$         &0.032&0.029&0.027  &$-$0.011&0.017&0.025  &$-$0.021&0.017&0.024  &$-$0.069&0.019&0.028  &$-$0.073&0.027&0.041\\
&$\lambda_\mathrm{inv}$ &$-$0.288&0.125&0.123  &$-$0.221&0.045&0.061  &$-$0.141&0.042&0.058  &$-$0.264&0.044&0.071  &$-$0.237&0.072&0.096\\
\midrule[1.0pt]
\multirow{4}{*}{5--7}&$\lambda_\theta$   &$-$0.265&0.040&0.062  &$-$0.147&0.014&0.024  &$-$0.095&0.015&0.023  &$-$0.029&0.019&0.030  &0.038&0.037&0.067\\
&$\lambda_{\theta\phi}$ &0.123&0.041&0.051  &$-$0.022&0.013&0.026  &$-$0.013&0.011&0.025  &0.026&0.013&0.028  &0.050&0.029&0.053\\
&$\lambda_\phi$         &$-$0.024&0.026&0.032  &0.033&0.014&0.024  &$-$0.024&0.015&0.024  &$-$0.060&0.018&0.030  &$-$0.125&0.029&0.051\\
&$\lambda_\mathrm{inv}$ &$-$0.330&0.080&0.098  &$-$0.049&0.040&0.059  &$-$0.163&0.035&0.056  &$-$0.198&0.040&0.067  &$-$0.299&0.066&0.106\\
\midrule[1.0pt]
\multirow{4}{*}{7--10}&$\lambda_\theta$   &0.035&0.039&0.044  &$-$0.078&0.019&0.028  &$-$0.098&0.020&0.031  &0.008&0.028&0.044  &0.225&0.061&0.082\\
&$\lambda_{\theta\phi}$ &0.006&0.038&0.046  &$-$0.002&0.014&0.023  &$-$0.034&0.013&0.021  &0.065&0.017&0.031  &$-$0.017&0.040&0.058\\
&$\lambda_\phi$         &$-$0.096&0.032&0.037  &$-$0.070&0.019&0.031  &$-$0.053&0.019&0.031  &$-$0.111&0.025&0.040  &$-$0.131&0.045&0.065\\
&$\lambda_\mathrm{inv}$ &$-$0.230&0.079&0.093  &$-$0.269&0.043&0.066  &$-$0.244&0.043&0.062  &$-$0.293&0.052&0.081  &$-$0.149&0.101&0.137\\
\midrule[1.0pt]
\multirow{4}{*}{10--15}&$\lambda_\theta$   &0.163&0.055&0.055  &0.042&0.037&0.042  &0.087&0.045&0.052  &0.138&0.063&0.089  &0.675&0.175&0.222\\
&$\lambda_{\theta\phi}$ &$-$0.103&0.043&0.065  &0.015&0.022&0.026  &$-$0.024&0.024&0.023  &0.062&0.034&0.046  &0.221&0.090&0.075\\
&$\lambda_\phi$         &$-$0.117&0.044&0.068  &$-$0.163&0.032&0.045  &$-$0.211&0.036&0.045  &$-$0.251&0.050&0.080  &$-$0.539&0.117&0.133\\
&$\lambda_\mathrm{inv}$ &$-$0.168&0.103&0.162  &$-$0.385&0.063&0.088  &$-$0.450&0.067&0.086  &$-$0.492&0.089&0.149  &$-$0.613&0.161&0.130\\
        \bottomrule[1.pt]

      \end{tabular}
    \end{center}
  \end{table}
\end{landscape}

\addcontentsline{toc}{section}{References}
\setboolean{inbibliography}{true}
\bibliographystyle{LHCb}
\bibliography{main,LHCb-PAPER,LHCb-CONF,LHCb-DP}

\end{document}